\documentclass[prb,twocolumn,showpacs,superscriptaddress]{revtex4-1}

\usepackage{graphicx}

\begin{document}
\title{Direct evidence of a zigzag spin chain structure in the honeycomb
lattice: A neutron and x-ray diffraction investigation of single crystal $\rm
Na_2IrO_3$}

\author{Feng Ye}
\author{Songxue Chi}
\author{Huibo Cao}
\author{Bryan~C.~Chakoumakos}
\affiliation{Quantum Condensed Matter Division, Oak Ridge National Laboratory,
Oak Ridge, Tennessee 37831, USA}
\author{Jaime~A.~Fernandez-Baca}
\affiliation{Quantum Condensed Matter Division, Oak Ridge National Laboratory,
Oak Ridge, Tennessee 37831, USA}
\affiliation{Department of Physics and Astronomy, University of Tennessee,
Knoxville, Tennessee 37996, USA}
\author{Radu Custelcean}
\affiliation{Chemical Science Division, Oak Ridge National Laboratory,
Oak Ridge, Tennessee 37831, USA}
\author{T.~F.~Qi}
\author{O.~B.~Korneta}
\author{G.~Cao}
\affiliation{Center for Advanced Materials, Department of Physics and
Astronomy, University of Kentucky, Lexington, Kentucky 40506, USA}
\date{\today}

\begin{abstract}
We have combined single crystal neutron and x-ray diffractions to investigate
the magnetic and crystal structures of the honeycomb lattice $\rm Na_2IrO_3$.
The system orders magnetically below $18.1(2)$~K with Ir$^{4+}$ ions forming
zigzag spin chains within the layered honeycomb network with an ordered moment of
$\rm 0.22(1)~\mu_B$/Ir site. Such a configuration sharply contrasts with the
N{\'{e}}el or stripe states proposed in the Kitaev-Heisenberg model. The
structure refinement reveals that the Ir atoms form a nearly ideal
two-dimensional honeycomb lattice while the $\rm IrO_6$ octahedra
experience a trigonal distortion that is critical to the ground
state.  The results of this study provide much needed experimental
insights into the magnetic and crystal structure that are crucial to the
understanding of the exotic magnetic order and possible topological
characteristics in the 5$d$-electron-based honeycomb lattice.
\end{abstract}

\pacs{75.25.-j,61.05.cf,75.50.Ee}

\maketitle

The 5$d$-based iridates have recently become a fertile yet largely
uncharted ground for studies of physics driven by the spin-orbit
coupling (SOC). It is now recognized that the SOC (0.4--1 eV), which
is proportional to $Z^4$ ($Z$ is the atomic number), plays a
critical role in the iridates, and rigorously competes with other
relevant energies, particularly the on-site Coulomb interaction $U$
(0.4 - 2.5 eV), which is significantly reduced because of the
extended nature of the 5$d$ orbitals.  A balance between the
competing energies is therefore established in the iridates and
drives exotic states seldom seen in other materials.  Recent
experimental observations and theoretical proposals for the iridates
have already captured the intriguing physics driven by SOC: $\rm
J_{eff} = 1/2$ Mott
states,\cite{kim08,moon08,kim09,ge11,chikara09,laguna10} spin
liquids in hyper-kagome structure,\cite{okamoto07} high-$T_C$
superconductivity,\cite{wang11} Weyl semimetals with Fermi
arcs,\cite{wan11a} correlated topological insulators with large
gaps,\cite{shitade09,pesin10} Kitaev model,\cite{jackeli09}
three-dimensional (3D) spin liquids with fermionic
spinons,\cite{zhou08} etc.

Of all iridates studied so far, $\rm Na_2IrO_3$ has inspired a great
deal of experimental and theoretical
efforts.\cite{singh10,liu11,singh12,pesin10,reuther11,yu12}  In
essence, the honeycomb lattice $\rm Na_2IrO_3$ is predicted to be a
topological insulator or a layered quantum spin Hall
insulator.\cite{shitade09}  However, conspicuous discrepancies among
various theoretical proposals and experimental observations clearly
point to the lack of a much needed characterization of the magnetic
and crystal structures of $\rm Na_2IrO_3$, whose band topology could
vary significantly with slight variations in the crystal structure.
This situation chiefly originates from the fact that the heavy
transition metals such as Ir strongly absorb neutrons, which makes a
comprehensive neutron study on the single crystal a nontrivial
challenge.

In this Rapid Communication, we report a combined neutron and x-ray
diffraction study on relatively large, thin single-crystal $\rm
Na_2IrO_3$.  This study reveals that Ir$^{4+}$ ions order
magnetically below 18.1(2)~K, and form zigzag spin chains along the
$a$ axis of the honeycomb structure with an ordered moment of
0.22(1) $\mu_B/$Ir.  Moreover, the structural refinements illustrate
that the Ir atoms feature a nearly perfect two-dimensional (2D)
honeycomb lattice and a trigonal distortion characterized by the
$\rm IrO_6$ octahedra deviating from a high-symmetric cubic
environment. These results are different from the previous x-ray
powder diffraction study, where the honeycomb was characterized by
three distinct bond lengths.\cite{singh10}

Single crystals of $\rm Na_2IrO_3$ were grown using a self-flux method from
off-stoichiometric quantities of IrO$_2$ and Na$_2$CO$_3$. Similar technical details
are described elsewhere.\cite{ge11,chikara09,laguna10} The crystals have
a circular basal area corresponding to the honeycomb plane with diameters of
$\sim$ 10~mm and thickness  $\sim$ 0.1~mm.  Such geometry provides a
unique advantage to significantly alleviate the technical difficulty
due to the inherent neutron absorption of the iridates.  Energy
dispersive x-ray spectroscopy (EDX) using a Hitachi/Oxford scanning
electron microscope (SEM)/EDX indicates a perfect stoichiometry of
$\rm Na_2IrO_3$ throughout the crystals studied.

The x-ray diffraction measurements were performed using a Bruker
SMART APEX CCD diffractometer with Mo K$_\alpha$ radiation and an
Oxford cryostream cooler. More than 40 crystals from four different
growth runs were screened at 125~K and full data sets were collected
on four crystals. (See Table I.) The neutron diffraction
measurements were carried out at the HB1A triple axis spectrometer
and HB3A four circle diffractometer at the High Flux Isotope Reactor
at the Oak Ridge National Laboratory with a fixed incident neutron
wavelength of $\rm \lambda=2.367$~and $\rm 1.536~\AA$,
respectively.  For the HB1A diffraction measurement, the crystal is
aligned in the $(0,k,l)$ scattering plane to allow the probing of
various magnetic reflections.

\begin{figure}[ht!]
\includegraphics[width=3.2in]{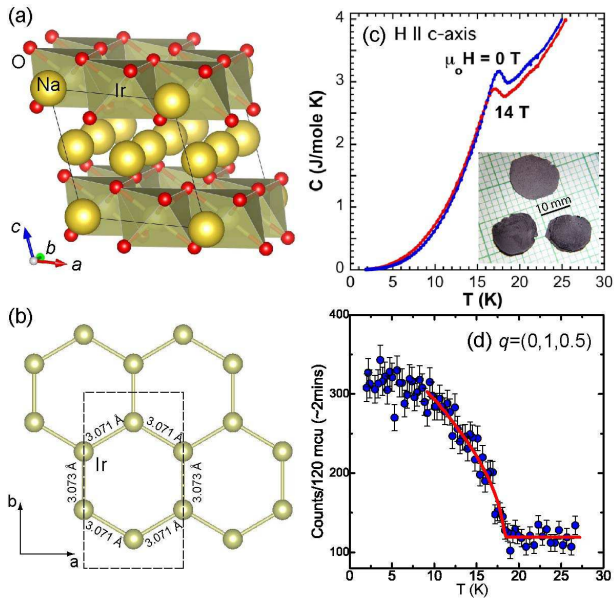}
\caption{(Color online) (a) Crystal structure of $\rm Na_2IrO_3$
with $C2/m$ symmetry.  (b) The honeycomb lattice formed by Ir atoms
within the basal plane with nearly equivalent distance between
neighboring Ir atoms. The dashed line denotes the unit cell. (c)
Specific heat $C(T)$ at $\rm H=0$ and $\rm H=14~T$.  Note that the
application of a magnetic field of 14~T suppresses the transition
temperature by only 0.5~K, apparently not characteristic of a
conventional N{\'{e}}el state. The inset shows the picture of single
crystals used for diffraction experiments. (d) The $T$ dependence of
the peak intensity of the $(0,1,0.5)$ magnetic reflection from
neutron diffraction measurement.  The solid line is the power law
fit described in the text.
}
\label{fig1}
\end{figure}

The systematic absences in the single-crystal x-ray diffraction
measurements unambiguously determine that the space group of $\rm
Na_2IrO_3$ is $C2/m$ and not $C2/c$ as initially
reported.\cite{singh10} This finding is consistent with a recent
single-crystal x-ray diffraction study by Choi {\it et
al.}\cite{choi12}  The typical crystal diffraction pattern shows
diffuse streaking, characteristic of stacking faults within the
layer sequence. Stacking faults involving fractional translation and
rotation of the fundamental $C2/m$ layer module have been modeled to
some extent in the iso-structural $\rm Li_2MnO_3$.\cite{breger05}
Polytypism analogous to that observed in the micas is also possible.
We adopted a structural model that allows for intermixing of the Na1
and Ir sites to artificially account for some amount of stacking
disorder, yet retain the ideal stoichiometry.  The overall structure
exhibits a virtually regular honeycomb layer of edge-sharing $\rm
IrO_6$ octahedra, similar to that observed in other so-called
dioctahedral sheets [e.g., gibbsite Al(OH)$_3$] in which the
octahedra are slightly flattened perpendicular to layer stacking.
In addition, the three O-Ir-O bond angles perpendicular to the basal
plane are all greater than 90$^\circ$ whereas the bond angles across
the shared edges are narrower, 84.1(3)$^\circ$ and 84.5(3)$^\circ$,
in contrast to the undistorted 90$^\circ$, as shown in Fig.~2(a).
The structural distortion indicates a presence of the trigonal
crystal field in addition to the cubic crystal field, due to the
repulsion of neighboring Ir atoms across the shared-edge of the
octahedra. The trigonal crystal field in $\rm Na_2IrO_3$ makes the
otherwise well separated gap between $\rm J_{eff}=1/2$ and $3/2$
levels\cite{kim09} less pronounced, highlighting an important role
in determining the electronic band structure topology.\cite{kim12}

\begin{table}[ht!]
\caption{Structural parameters at $T$=125~K from single crystal x-ray
diffraction measurements. The full data sets could be indexed using space
group $C2/m$ with  $a=5.319(1)~\rm \AA$, $b=9.215(2)~\rm \AA$, $c=5.536(1)~\rm
\AA$, and $\beta=108.67(1)^{\circ}$.  The Ir-O bond distances are 2.069(8),
2.067(9), and 2.060(12)~$\rm \AA$, and the Ir$\cdots$Ir distances are 3.073(1)
and 3.0705(8)~$\rm \AA$.  Refinements are made using
SHELXL-97 (Ref.~\onlinecite{shelx08}), yielding an agreement factor
$R1=0.0687$ for 334 reflections with $F_{obs}>4\sigma(F_{obs})$.
}
\label{tab1}
\begin{ruledtabular}
\begin{tabular}{llccccc}
 & Site & $x$  & $y$  & $z$ & Occupancy & $U(\AA^2)$ \\
    \hline
  Ir1 & 4g & 0  	& 0.3332(1) 	& 0 	& 0.823(6) & 0.006(1)  \\
  Na4 & 4g & 0  	& 0.3332(1) 	& 0 	& 0.177(6) & 0.006(1)  \\
  Na1 & 2a & 0  	& 0         	& 0 	& 0.646(9) & 0.014(2)  \\
  Ir2 & 2a & 0  	& 0         	& 0 	& 0.354(9) & 0.014(2)  \\
  Na2 & 4h & 0  	& 0.8363    	& 1/2 	& 1 & 0.003(2) \\
  Na3 & 2d & 0  	& 1/2       	& 1/2 	& 1 & 0.004(2)  \\
  O1 & 8j  & 0.259(3)  	& 0.3294(7) 	& 0.792(3) & 1 & 0.001(3) \\
  O2 & 4i  & 0.270(3)  	& 0 		& 0.792(3) & 1 & 0.001(3) \\
  \end{tabular}
\end{ruledtabular}
\end{table}

\begin{figure}[ht!]
\includegraphics[width=3.2in]{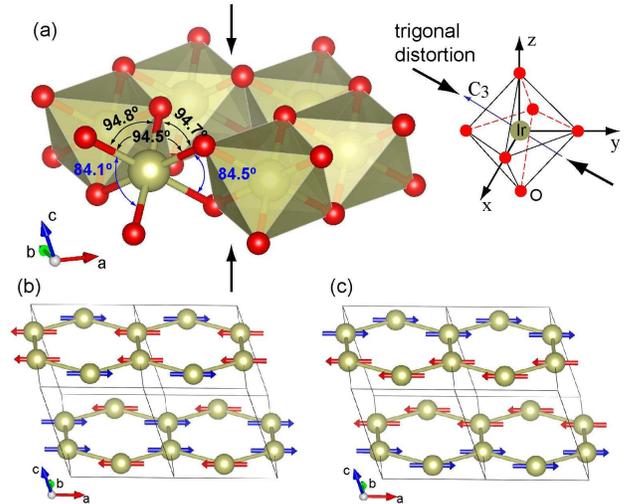}
\caption{(Color online) (a) Local structure within the basal plane. The
compression of $\rm IrO_6$ octahedron along the stacking
leads to the decrease of O-Ir-O bond angles across the shared edges. (b)-(c)
Comparison of stripe and zigzag order that are consistent with the symmetry
associated with observed magnetic reflections. In both cases, the Ir moments
between honeycomb layers are antiferromagnetically coupled.
}
\label{fig2}
\end{figure}

The magnetic ground state is further characterized by the neutron
diffraction on the single crystals. The magnetic propagation wave
vector was determined to be ${\bf q}_m=(0,1,0.5)$ in the $C2/m$
notation based on an extensive survey in reciprocal space using the
four-circle neutron diffractometer.  Figure 1(d) shows that the
magnetic Bragg peak intensity ($I_B\propto |M_s|^2$, $M_s$ is the
order parameter) disappears above $T_N=18.1\pm0.2$~K, consistent
with the anomaly observed in the specific heat data [Fig.~1(c)].
Fitting $I_B$ to the power law scaling function of
$(1-T/T_N)^{2\beta}$ yields a critical exponent $\beta=0.29(2)$ that
is typical of a three-dimensional magnetic system.  The
determination of a magnetic propagation wave vector and the correct
description of the crystal structure put stringent constraints on
the possible magnetic models. The magnetic reflection appearing at
(0,1,0.5) rules out the N{\'{e}}el configuration [characterized by
antiferromagnetically-coupled nearest neighboring spins with ${\bf
q}'_m=(0,0,0.5)$] but leaves the choice of either stripe or zigzag
order in the basal plane as depicted in Figs.~2(b) and 2(c).  Group
theory analysis indicates that the magnetic representation $\rm
\Gamma_{mag}$ can be decomposed into an irreducible representation
(IR) $\rm \Gamma_{mag}=\Gamma_1+\Gamma_2+2\Gamma_3+2\Gamma_4$ with
corresponding basis vectors (BVs) listed in Table~II. Since the
moment direction has been characterized to be along the $a$ axis by
magnetic susceptibility and polarized x-ray
measurements,\cite{liu11} this information is implemented to perform
the model calculation and magnetic structural refinement.  Figures
3(a)-3(c) show the rocking scans of three characteristic magnetic
Bragg reflections ${\bf q}_1=(0,1,0.5)$, ${\bf q}_2=(0,3,1.5)$, and
${\bf q}_3=(0,3,0.5)$ in the $(0,k,l)$ scattering plane.  The
strongest reflection occurs at ${\bf q}_1$ and the intensity
decrease sharply at ${\bf q}_2$ that has a larger momentum transfer.
In contrast, there is no sign of magnetic scattering at ${\bf q}_3$
at base temperature. For single-crystal magnetic scattering at
wave vector transfer ${\bf q}$, the measured intensity follows
\begin{equation}	
 |F_{\perp}({\bf q})|^2 = |{\bf F}_m({\bf q})|^2 - [\hat{\bf e} \cdot {\bf F}_m({\bf q})]^2,
\end{equation}
where $\hat{\bf e}$ is the unit vector along the ${\bf q}$, and ${\bf
F}_{m}(\bf {q})$ is the magnetic structure factor that can be expressed as
\begin{equation}
{\bf F}_m({\bf q}) = p \sum_{j=1}^{n} f_j({\bf q}) {\bf S}_{{\bf k},j} \exp
{2\pi i ({\bf q} \cdot {\bf r}_j}).
\end{equation}
Here the sum is over all the magnetic atoms in the crystallographic cell,
$p=r_e \gamma/2=0.2695$, ${\bf S}_{{\bf k},j}$ are the Fourier components
proportional to the BVs listed in Table~II, ${\bf r}$ is the vector position
of atom $j$, and $f({\bf q})$ is the magnetic form factor for the Ir$^{4+}$
ions.\cite{Irformfactor}

\begin{table}[ht!]
\caption{Basis vectors (BVs) $\psi_i$ of an IR of the space group $C2/m$ and ${\bf
k}=(0,1,0.5)$. BVs are defined relative to the crystallographic axes.
Magnetic moments for $j$ atom in $l^{\rm th}$ cell are given by ${\bf
m}_{l,j}=\sum_{\bf k} {\bf S}_{{\bf k},j} \exp(-2\pi i \bf{k} \cdot {\bf
R}_l)$ and ${\bf S}_{{\bf k},j}=\sum_i C_i \psi_i$, where $C_i$ is the mixing
coefficient. Only $\Gamma_3$ and $\Gamma_4$ are relevant since they describe
the correct spin direction along the $a$ axis.}
\label{tab2}
\begin{ruledtabular}
\begin{tabular}{lcccc}
 & $\psi_1(\Gamma_1)$ & $\psi_2(\Gamma_2)$  & $\psi_3,\psi_4(\Gamma_3)$  & $\psi_5,\psi_6(\Gamma_4)$ \\
    \hline
  Ir~$(0,0.333,0)$  & (0,1,0) & (0,1,0) & (1,0,0),(0,0,1) & (1,0,0),(0,0,1)  \\
  Ir~$(0,0.667,0)$  & (0,1,0) & (0,-1,0) & (1,0,0),(0,0,1) & (-1,0,0),(0,0,-1)  \\
  \end{tabular}
\end{ruledtabular}
\end{table}

\begin{table}[ht!]
\caption{Calculated magnetic scattering $|F_{\perp}({\bf q}_i)|^2$
at ${\bf q}_1$ (normalized to 100), ${\bf q}_2$, and ${\bf q}_3$ for
stripe and zigzag spin orders, and their comparison to the
measurement.  The errorbar is statistical and refers to one standard
deviation.
}
\label{tab3}
\begin{ruledtabular}
\begin{tabular}{lcccc}
 & ${\bf q}_1=(0, 1, 0.5)$ & ${\bf q}_2=(0, 1, 1.5)$  & ${\bf q}_3=(0, 3, 0.5)$ \\
    \hline
  $\Gamma_3$ (stripe) & 100 & 51.1 & 186.5   \\
  $\Gamma_4$ (zigzag) & 100 & 51.1 & 0.002  \\
  Measurement & $5.40\pm0.38$ & $2.77\pm0.32$ & 0  \\
  \end{tabular}
\end{ruledtabular}
\end{table}

As summarized in Table~III, both stripe and zigzag spin orders give
the identical ratio $|F_{\perp}({\bf q}_2)/F_{\perp}({\bf q}_1)|^2$.
Therefore the magnetic scattering at these two reflections alone
cannot distinguish the difference between the two spin
configurations. However, the magnetic scattering at ${\bf q}_3$ is
expected to be strong for the stripe spin configuration but absent
for the zigzag spin chains; the absence of the magnetic scattering
at ${\bf q}_3$ illustrated in Fig.~3(c) clearly indicates a presence
of the zigzag spin order. To determine the magnitude of the magnetic
moment/Ir, a large set of nuclear reflections under the same
experimental configuration were collected to get the scale factor
for normalization, yielding a magnetic moment of $\rm
0.22(1)~\mu_B/Ir$; this is considerably smaller than that ($\rm
1~\mu_B$/Ir) for an $S=1/2$ system, consistent with early
observations for systems such as $\rm Sr_2IrO_4$ and $\rm BaIrO_3$
where the ordered moment is no more than 15\% of $\rm
1~\mu_B$/Ir.\cite{ge11,chikara09,laguna10,cao98} The significantly
reduced moment might be ascribed to the strong hybridization of the
Ir $5d$ orbital with the ligand oxygen $2p$ orbital and the moments
are largely canceled out in the antiferromagnetic state.  Moreover,
the wave vector scans presented in Figs.~3(d) and 3(e) show a
resolution limited Gaussian profile and Lorentzian-like lineshape
for the in-plane and out-of-plane magnetic correlation,
respectively. The data reinforce that the spins form a long-range
order in the honeycomb basal plane while a short range order might
still remain to some extent (with a correlation length $\xi \approx
\rm 139\pm21~\AA$) between layers due to the inherent imperfection
in crystal structure, as suggested in the x-ray diffraction presented
above.

\begin{figure}[ht!]
\includegraphics[width=3.2in]{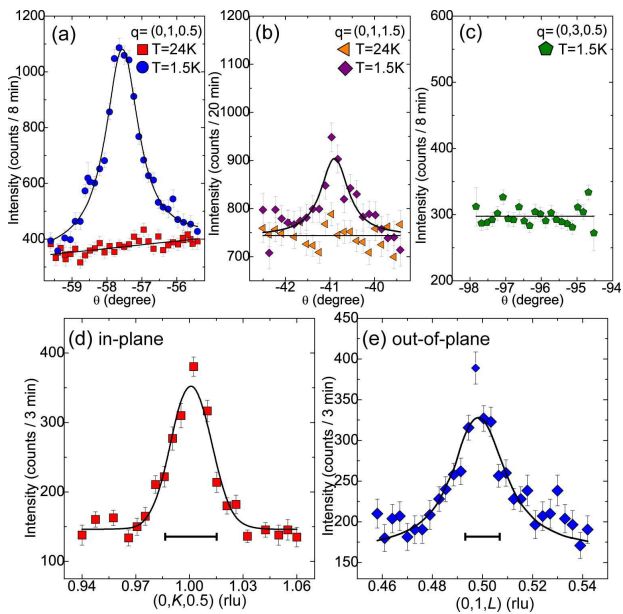}
\caption{(Color online) The rocking scans of characteristic magnetic
reflections of (a) $(0,1,0.5)$, (b) $(0,1,1.5)$, and (c)
$(0,3,0.5)$.  (d), (e) The in-plane and out-of-plane wave vector
scans for the $(0,1,0.5)$ peak. The solid line in (e) is the fit to
the Lorentzian form with instrument resolution convoluted. The
horizontal bars in (d) and (e) denote the instrument resolution.
}
\label{fig3}
\end{figure}

In recent theoretical proposals, $\rm Na_2IrO_3$ is regarded as one
of a few model systems that can be mapped into the exactly solvable
Kitaev model.  \cite{kitaev06} The combination of isotropic
Heisenberg exchange interaction and anisotropic Kitaev term through
strong spin-lattice coupling gives rise to a rich variety of low
energy magnetic ground states.  This includes the topologically
nontrivial quantum spin Hall system in the weak interaction limit
\cite{shitade09} and evolution from the conventional N{\'e}el order
to the spin liquid state sandwiched by a stripe phase depending on
the microscopic parameters in the strong spin-orbit coupling limit.
\cite{jackeli09,chaloupka10} The geometric frustration due to the
longer range exchange paths and the dynamic frustration caused by
the Kitaev term leave the physical properties of $\rm Na_2IrO_3$
highly tunable by small perturbations, such as magnetic field,
vacancies, and structural distortions.
\cite{kimchi09,jiang11,trousselet11,you11} Only recently has the
magnetic ground state been experimentally examined and proposed to
be a possible zigzag spin state using resonant magnetic x-ray
scattering.  \cite{liu11} The unexpected spin state inconsistent
with the original Kitaev-Heisenberg model underscores the novelty of
the magnetic ground state, prompting theoretical suggestions that
the zigzag magnetic order could be explained only when the
long-range magnetic Heisenberg interactions ($J_2,
J_3$)\cite{kimchi09,singh12} or a trigonal distortion of the $\rm
IrO_6$ octahedra \cite{bhattacharjee11,yang10} in the $[1,1,1]$
direction (local basis of the octahedron) is taken into account.
Indeed, a unique quantum phase transition from normal to topological
insulator is recently predicted in $\rm Na_2IrO_3$ if
both the long-range hopping and trigonal crystal field terms are
included.  \cite{kim12} On the other hand, noticeable
inconsistencies still exist in the band topology predictions that
are likely due to the structural parameters used for the
first-principles calculations.\cite{shitade09,jin09} With the
presence of the trigonal crystal field, it is suggested that the
$\rm J_{eff}=1/2$ doublet is no longer as critical in $\rm
Na_2IrO_3$ as in $\rm
Sr_2IrO_4$,\cite{kim08,moon08,kim09,ge11,chikara09} $\rm BaIrO_3$,
\cite{laguna10} and other layered iridates; instead, the trigonal
crystal field (0.6~eV) and long-range hopping dictate the topological
character, which is extremely sensitive to slight structural
changes.\cite{kim12}

One of the unique aspects of this work is that both neutron and x-ray
diffraction data were collected from single crystals of $\rm Na_2IrO_3$.  The
results of this work therefore provide the well-defined characteristics of the
magnetic and crystal structures of the honeycomb lattice, and significantly
improve our understanding of this intriguing system. We expect this study will
help clarify the topological character of the ground state in $\rm Na_2IrO_3$,
a fertile ground yet to be fully explored.

We thank S. Okamoto and C. de la Cruz for invaluable discussions.
Research at ORNL was sponsored in part by the Division of Chemical
Sciences, Geosciences, and Biosciences, and the Scientific User
Facilities Division, Office of Basic Energy Sciences, U.S.
Department of Energy.  The work at University of Kentucky was
supported by NSF through Grants No.~DMR-0856234 and No.~EPS-0814194.


\begin{thebibliography}{10}
\bibitem{kim08}
B.~J. Kim {\em et~al.},
\newblock Phys.\ Rev.\ Lett. {\bf 101}, 076402 (2008).

\bibitem{moon08}
S.~J. Moon {\em et~al.},
\newblock Phys.\ Rev.\ Lett. {\bf 101}, 226402 (2008).

\bibitem{kim09}
B.~J. Kim {\em et~al.},
\newblock Science {\bf 323}, 1329 (2009).

\bibitem{ge11}
M.~Ge {\em et~al.},
\newblock Phys.\ Rev.\ B {\bf 84}, 100402(R) (2011).

\bibitem{chikara09}
S.~Chikara {\em et~al.},
\newblock Phys.\ Rev.\ B {\bf 80}, 140407(R) (2009).

\bibitem{laguna10}
M.~A. Laguna-Marco {\em et~al.},
\newblock Phys.\ Rev.\ Lett. {\bf 105}, 216407 (2010).

\bibitem{okamoto07}
Y.~Okamoto, M.~Nohara, H.~Aruga-Katori, and H.~Takagi,
\newblock Phys.\ Rev.\ Lett. {\bf 99}, 137207 (2007).

\bibitem{wang11}
F.~Wang and T.~Senthil,
\newblock Phys.\ Rev.\ Lett. {\bf 106}, 136402 (2011).

\bibitem{wan11a}
X.~Wan, A.~M. Turner, A.~Vishwanath, and S.~Y. Savrasov,
\newblock Phys.\ Rev.\ B {\bf 83}, 205101 (2011).

\bibitem{shitade09}
A.~Shitade {\em et~al.},
\newblock Phys.\ Rev.\ Lett. {\bf 102}, 256403 (2009).

\bibitem{pesin10}
D.~A. Pesin and L.~Balents,
\newblock Nat.\ Phys. {\bf 6}, 376 (2010).

\bibitem{jackeli09}
G.~Jackeli and G.~Khaliullin,
\newblock Phys.\ Rev.\ Lett. {\bf 102}, 017205 (2009).

\bibitem{zhou08}
Y.~Zhou, P.~A. Lee, T.-K. Ng, and F.-C. Zhang,
\newblock Phys.\ Rev.\ Lett. {\bf 101}, 197201 (2008).

\bibitem{singh10}
Y.~Singh and P.~Gegenwart,
\newblock Phys.\ Rev.\ B {\bf 82}, 064412 (2010).

\bibitem{liu11}
X.~Liu {\em et~al.},
\newblock Phys.\ Rev.\ B {\bf 83}, 220403(R) (2011).

\bibitem{singh12}
Y.~Singh {\em et~al.},
\newblock Phys.\ Rev.\ Lett. {\bf 108}, 127203 (2012).

\bibitem{reuther11}
J.~Reuther, R.~Thomale, and S.~Trebst,
\newblock Phys. Rev. B {\bf 84}, 100406(R) (2011).

\bibitem{yu12}
Y.~Yu and S.~Qin,
\newblock arXiv:1202.1610  (2012).

\bibitem{choi12}
S.~Choi {\em et~al.},
\newblock Phys. Rev. Lett. {\bf 108}, 127204 (2012).

\bibitem{breger05}
J.~Br\'{e}ger {\em et~al.},
\newblock Journal of Solid Chemistry {\bf 178}, 2575 (2008).

\bibitem{kim12}
C.~H. Kim, H.~S. Kim, H.~Jeong, H.~Jin, and J.~Yu,
\newblock Phys. Rev. Lett. {\bf 108}, 106401 (2012).

\bibitem{shelx08}
G.~M. Sheldrick,
\newblock Acta Cryst. {\bf A64}, 112 (2008).

\bibitem{Irformfactor}
K.~Kobayashi, T.~Nagao, and M.~Ito,
\newblock Acta Cryst A {\bf 67}, 473 (2011).

\bibitem{cao98}
G.~Cao, J.~Bolivar, S.~McCall, J.~E. Crow, and R.~P. Guertin,
\newblock Phys.\ Rev.\ B {\bf 57}, R11039 (1998).

\bibitem{kitaev06}
A.~Kitaev,
\newblock Ann. Phys. {\bf 321}, 2 (2006).

\bibitem{chaloupka10}
J.~Chaloupka, G.~Jackeli, and G.~Khaliullin,
\newblock Phys.\ Rev.\ Lett. {\bf 105}, 027204 (2010).

\bibitem{kimchi09}
I.~Kimchi and Y.~Z. You,
\newblock Phys.\ Rev.\ B {\bf 84}, 180407(R) (2011).

\bibitem{jiang11}
H.~C. Jiang, Z.~C. Gu, X.~L. Qi, and S.~Trebst,
\newblock Phys.\ Rev.\ B {\bf 83}, 245104 (2011).

\bibitem{trousselet11}
F.~Trousselet, G.~Khaliullin, and P.~Horsch,
\newblock Phys.\ Rev.\ B {\bf 84}, 054409 (2011).

\bibitem{you11}
Y.~Z. You, I.~Kimchi, and A.~Vishwanath,
\newblock arXiv:1109.4155  (2011).

\bibitem{bhattacharjee11}
S.~Bhattacharjee, S.~Lee, and Y.~B. Kim,
\newblock arXiv:1108.1806  (2011).

\bibitem{yang10}
B.~J. Yang and Y.~B. Kim,
\newblock Phys.\ Rev.\ B {\bf 82}, 085111 (2010).

\bibitem{jin09}
H.~Jin, H.~Kim, H.~Jeong, C.~H. Kim, and J.~Yu,
\newblock arXiv:0907.0743  (2009).

\end{thebibliography}
\end{document}